\documentclass[preprint2,tighten,times,iop]{emulateapj}

\usepackage[utf8]{inputenc}
\usepackage{mathtools}
\usepackage{graphicx}
\usepackage{float}
\usepackage{multirow}
\usepackage[english]{babel}
\usepackage{blindtext}
\usepackage{hyperref}
\usepackage{xcolor}
\hypersetup{
    colorlinks,
    linkcolor={red!50!black},
    citecolor={blue!50!black},
    urlcolor={blue!80!black}
}

\begin{document}

\title{A Dipole on the Sky : Predictions for Hypervelocity Stars from the Large Magellanic Cloud}

\shorttitle{A Dipole on the Sky}
\shortauthors{Boubert and Evans}
\author{Douglas Boubert}
\author{N. Wyn Evans}
\affil{Institute of Astronomy, University of Cambridge, Madingley Road, Cambridge CB3 0HA, UK; \href{mailto:d.boubert@ast.cam.ac.uk}{d.boubert@ast.cam.ac.uk}, \href{mailto:nwe@ast.cam.ac.uk}{nwe@ast.cam.ac.uk}}

\begin{abstract}
We predict the distribution of hypervelocity stars (HVSs) ejected from
the Large Magellanic Cloud (LMC), under the assumption that the dwarf
galaxy hosts a central massive black hole (MBH). For the majority of
stars ejected from the LMC the orbital velocity of the LMC has
contributed a significant fraction of their galactic rest frame
velocity, leading to a dipole density distribution on the sky. We
quantify the dipole using spherical harmonic analysis and contrast
with the monopole expected for HVSs ejected from the Galactic Center
(GC). There is a tendril in the density distribution that leads the
LMC which is coincident with the well-known and unexplained clustering
of HVSs in the constellations of Leo and Sextans. Our model is
falsifiable, since it predicts that Gaia will reveal a large density
of HVSs in the southern hemisphere.
\end{abstract}

\maketitle
\begin{figure*}[t!]
\includegraphics[scale=1.0,trim = 9mm 30mm 15mm 35mm, clip]{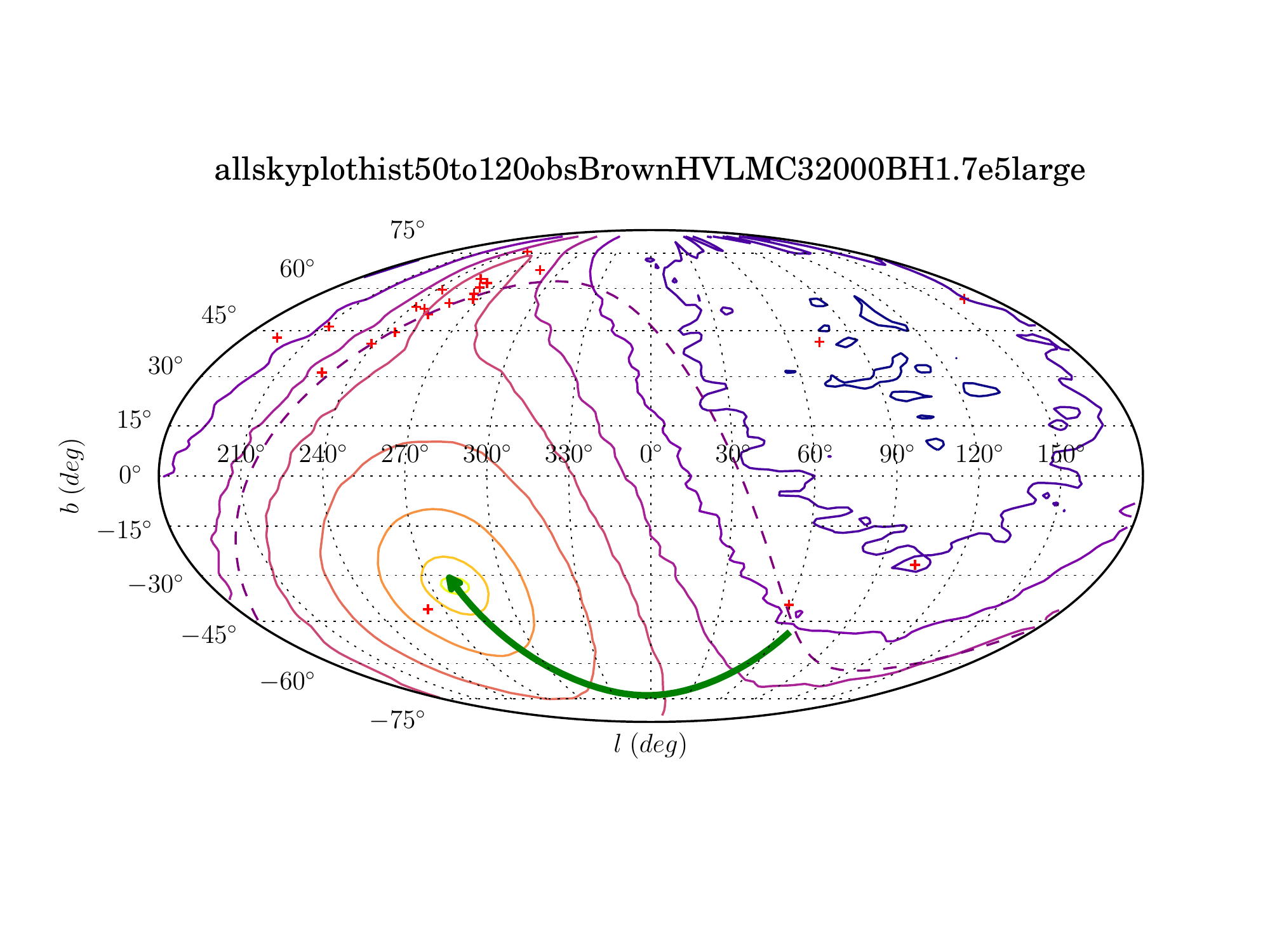}
\caption{Predicted distribution in galactic coordinates of HVSs from 
  the LMC at a heliocentric distance $50<d<120 \; \mathrm{kpc}$, using 
  the HVLMC-A model from Section \ref{sec:method}. Density contours are spaced $10^{1/2}$
  apart with the yellow contour denoting the highest density, the green arrow marks the path of the LMC over the last $350
  \; \mathrm{Myr}$, the purple dashed line is the celestial equator, and the
  red crosses are the HVS candidates denoted HVS1, HVS2, etc.  by
  Brown.}
\label{fig:feature}
\end{figure*}

\section{Introduction}

A \emph{hypervelocity star} (HVS) is any star whose velocity in the
galactic rest frame exceeds the local escape speed of the Milky Way
(MW). Several routes have been suggested for their production, the
first being reported by \citet{hills_hyper-velocity_1988} who
theorised that the tidal disruption of a stellar binary during a close
encounter with the massive black hole (MBH) at the centre of the MW
could accelerate one member of the binary, in what is now known as the
Hills mechanism.

The first candidate HVS, a B-type star travelling with a galactic rest
frame velocity of $709 \; \mathrm{km} \; \mathrm{s}^{-1}$, was
discovered by \citet{brown_discovery_2005}. Subsequent surveys have
discovered dozens of further candidates. The recent review paper
\citet{brown_hypervelocity_2015} gives a summary, whilst there are
subsequent papers by \citet{theissen_warm_2014,savcheva_new_2014,
  hawkins_characterizing_2015,kunder_high-velocity_2015,
  li_19_2015,vickers_red_2015,favia_runaway_2015}. It was noticed
early on that the candidate HVSs exhibited significant spatial
anisotropy on the sky with 8 out of the 14 HVSs at that time being
located in two constellations despite a fifth of the sky having been
surveyed \citep{brown_anisotropic_2009}.

It is a natural extension of the Hills mechanism to ask where else it may
occur. For instance \citet{lu_hypervelocity_2007} and
\citet{sherwin_hypervelocity_2008} suggest the central MBHs of M31 and
M32 as possible sites. The recent discovery of a red supergiant
runaway candidate of M31 by \citet{evans_runaway_2015} suggests these
two possibilities may soon be testable. The Large Magellanic Cloud
(LMC) is the most massive of the MW's dwarf satellites and thus there
is the potential for the LMC to host a central MBH. The Hills mechanism
could accelerate large numbers of stars to a few $100 \; \mathrm{km}
\; \mathrm{s}^{-1}$ above the escape speed from the LMC. The LMC has
an orbital velocity of $378 \; \mathrm{km} \; \mathrm{s}^{-1}$
\citep{van_der_marel_third-epoch_2014}, thus alignment of these
velocity vectors could result in HVSs that have been ejected from the
LMC and are travelling at the escape speed of the MW. The location of
the LMC on the sky and orientation of its velocity vector could
explain the spatial anisotropy in the observed HVSs.

\section{Method}
\label{sec:method}

To test the hypothesis that there is a significant number of HVSs
ejected from the LMC, we require a prediction for the observables of
such a population and a reference population of HVSs from the Galactic
Center (GC). We consider three models; HVGC is a model that produces stars
from the GC and is identical to HV3 from
\citet{kenyon_predicted_2014}, HVLMC-A is a population from the LMC,
and HVLMC-B is identical to HVLMC-A but has the initial ejection
velocity cut by $200 \; \mathrm{km} \; \mathrm{s}^{-1}$ to approximate
the escape velocity from the centre of the LMC. All three models
consider $3 \; \mathrm{M}_{\odot}$ HVSs, which have a typical main sequence 
lifetime of $350 \; \mathrm{Myr}$, since there is a large sample
of B-type stars in the HVS literature we can compare to
\citep{brown_hypervelocity_2015}. We simulate the process of ejection
from the centre of the LMC by seeding HVSs along the orbit of the LMC
over the past $350 \; \mathrm{Myr}$, with a velocity which is the sum
of the orbital velocity of the LMC and the ejection velocity due to
the Hills mechanism. These stars are then evolved through a potential
model of the MW until a randomly selected observation time, at which
point we record the position and velocity in the galactic rest
frame. To allow for cross-comparisons with
\citet{kenyon_predicted_2014}, we assume that the Sun is at a position
$(x,y,z)=(-R_{\odot},0,0)$ relative to the GC where
$R_{\odot}=8 \; \mathrm{kpc}$, and is travelling with a velocity
$(v_x,v_y,v_z)=(0,v_{\odot},0)$ where $v_{\odot}=235 \; \mathrm{km} \;
\mathrm{s}^{-1}$. 

%
%

\subsection{LMC black hole mass}
\label{sec:bh}

Observations of dwarf galaxy hosts of MBHs have recently been carried
out by \citet{reines_dwarf_2013}, who specifically reference these
galaxies as being in the mass range of the LMC. For dwarf galaxies
with stellar masses in the range $10^{8.5} \lesssim M_{\mathrm{star}}
\lesssim 10^{9.5} \; M_{\odot}$ they find MBH masses of $10^5 \lesssim
M_{\mathrm{bh}} \lesssim 10^{6} \; M_{\odot}$, with a median of
$M_{\mathrm{bh}} \sim 2\times10^{5} \; M_{\odot}$. Since the
dependence of the mean of the ejection velocity from the
Hills mechanism is only weakly dependent on the MBH mass
($\propto M_{\mathrm{bh}}^{1/6}$), see Section \ref{sec:hills}, a
factor of 10 in MBH mass is only a factor of $\sim 1.5$ in the mean
ejection velocity. Using Equation \ref{eq:vejH} from Section
\ref{sec:hills}, we see that the mean ejection velocity for the
binaries with a separation of $0.115 \; \mathrm{AU}$ is $\approx 1800\;
\mathrm{km}\;\mathrm{s}^{-1}$, thus MBHs which are $10^4$ times less
massive are still capable of ejecting HVSs at velocities of $\geq 350\;
\mathrm{km} \; \mathrm{s}^{-1}$. We thus assume a fiducial value
$M_{\mathrm{bh}}=1.7\times10^5 \; M_{\odot}$, similar to the median of
\citet{reines_dwarf_2013}.

\subsection{Hills mechanism}
\label{sec:hills}

We first require the initial positions $\boldsymbol{\widetilde{r}}_0$
and velocities $\boldsymbol{\widetilde{v}}_0$ of the candidate HVSs in
a frame co-moving with the MBH from which they were ejected. This
section uses the prescription of \citet{kenyon_predicted_2014}, itself based
on the Hills mechanism algorithm of \citet{bromley_hypervelocity_2006},
which we briefly describe here to aid clarity. At the time of their 
ejection the stars are uniformly distributed across the surface of a sphere 
with radius $1.4\;\mathrm{pc}$ centered on the MBH and given an initial 
outwardly radial trajectory. Thus the only free parameters not yet 
determined are the magnitude of the velocity $v_{\mathrm{ej}}$, 
the age of the star when it is ejected $t_{\mathrm{ej}}$, and the age
 of the star when it is observed $t_{\mathrm{obs}}$.

Suppose that an equal mass binary with separation $a_{\mathrm{bin}}$
interacts at a closest approach $r_{\mathrm{close}}$ with a MBH of
mass $M_{\mathrm{bh}}$. One star of the system will be ejected by the Hills mechanism with probability
\begin{equation}
\label{eq:pej}
P_{\mathrm{ej}}=1-D/175
\end{equation}
where
\begin{equation}
\label{eq:d}
D=D_0\left(\frac{r_{\mathrm{close}}}{a_{\mathrm{bin}}}\right)\; , \; D_0=\left[\frac{2M_{\mathrm{bh}}}{10^6(M_1+M_2)}\right]^{-1/3}.
\end{equation}
The probability $P$ that a star is ejected is drawn from a uniform
distribution between 0 and 1. If the star is ejected, then following
\citet{kenyon_predicted_2014}, the velocity is drawn from the
distribution
\begin{equation}
\label{eq:pvej}
P(v_{\text{ej}})\text{d}v_{\text{ej}} \propto \exp{\left(-\frac{( v_{\text{ej}} - v_{\text{ej,H}} )^2}{2\sigma_v^2}\right)} \text{d} v_{\text{ej}},
\end{equation}
where $\sigma_v=0.2 v_{\text{ej,H}}$ and $v_{\text{ej,H}}$ is given by
\begin{align}
\label{eq:vejH}
v_{\mathrm{ej,H}}=&1760\left(\frac{a_{\mathrm{bin}}}{0.1 \; \mathrm{AU}}\right)^{-1/2}\left(\frac{M_1+M_2}{2 \; \mathrm{M}_{\odot}}\right)^{1/3}\nonumber  \\
                  &\times\left(\frac{M_{\mathrm{bh}}}{3.5 \times 10^6 \; \mathrm{M}_{\odot}}\right)^{1/6} f_{\mathrm{R}}\; \mathrm{km}\;\mathrm{s}^{-1}.
\end{align}
We only consider equal mass binaries so $M_1=M_2$. Note that
$f_{\mathrm{R}}$ is a normalisation factor \citep{kenyon_predicted_2014}
\begin{align}
\label{eq:fr}
f_{\mathrm{R}}&=0.774+(0.0204+(-6.23 \times 10^{-4} + (7.62\times 10^{-6} \nonumber \\
               &+(-4.24\times 10^{-8}+8.62\times 10^{-11}D) D)D)D)D.
\end{align}
Since we are interested in all possible equal mass binary systems that
can result in an ejection we sample from distributions of
$a_{\mathrm{bin}}$ and $r_{\mathrm{close}}$. The separation of the
binary $a_{\text{bin}}$ is drawn from
\begin{equation}
P(a_{\text{bin}})\text{d}a_{\text{bin}}\propto \frac{1}{a_{\text{bin}}}\text{d}a_{\text{bin}},
\end{equation}
where $a_{\text{bin,min}}= 0.115\text{ AU}$ for stars of mass $ 3 \text{ M}_{\odot}$ and $a_{\text{bin,max}}= 4\text{ AU}$. The closest approach $r_{\text{close}}$ has 
\begin{equation}
P(r_{\text{close}})\text{d}r_{\text{close}}\propto r_{\text{close}}\text{d}r_{\text{close}},
\end{equation}
where $r_{\text{close,min}}= 1 \text{ AU}$ and $r_{\text{close,max}}= 175 a_{\text{bin,max}}/D_0$. 

Both the age of the star at the time of ejection $t_{\text{ej}}$ and
observation $t_{\text{obs}}$ are drawn from the uniform distribution
\begin{equation}
P(t)\text{d}t \propto \text{d} t,
\end{equation}
where $t_{\text{min}}= 0 \text{ Gyr}$ and $t_{\text{max}}=
t_{\text{ms}}$ and the main sequence lifetime $t_{\text{ms}} =
0.35\text{ Gyr}$ for stars of mass $3 \text{ M}_{\odot}$. The time of
flight of the star from ejection to observation is then defined by
$t_{\mathrm{f}}=t_{\mathrm{obs}}-t_{\mathrm{ej}}$.

We discard any sampled star that does not satisfy $P_{\text{ej}}\geq P$,
$v_{\text{ej}}\geq v_{\text{ej,min}}$ and
$t_{\text{ej}}<t_{\text{obs}}$. Note that $v_{\text{ej,min}}$ is
chosen to minimise the computation time spent on stars that will be
difficult to observe. For the HVGC model this corresponds to stars
that will remain in the GC so $v_{\text{ej,min}}=600 \;
\mathrm{km} \; \mathrm{s}^{-1}$. For stars in the HVLMC-A and HVLMC-B
models, we choose $v_{\text{ej,min}}=200 \; \mathrm{km} \;
\mathrm{s}^{-1}$ to ensure we only include stars that escape the LMC.

\subsection{Large Magellanic Cloud orbit}
\label{sec:lmcorbit}

For stars ejected from the centre of the MW, the position
$\boldsymbol{r}_0$ and velocity $\boldsymbol{v}_0$ vectors in the
galactic rest frame satisfy
$\boldsymbol{r}_0=\boldsymbol{\widetilde{r}}_0$ and
$\boldsymbol{v}_0=\boldsymbol{\widetilde{v}}_0$. However, for stars
ejected from the LMC
\begin{equation}
\boldsymbol{r}_0=\boldsymbol{\widetilde{r}}_0 + \boldsymbol{r}_{\mathrm{LMC}}(t_{\mathrm{f}}) \; , \; \boldsymbol{v}_0=\boldsymbol{\widetilde{v}}_0 + \boldsymbol{v}_{\mathrm{LMC}}(t_{\mathrm{f}}),
\end{equation}
where $\boldsymbol{r}_{\mathrm{LMC}}(t_{\mathrm{f}})$ and
$\boldsymbol{v}_{\mathrm{LMC}}(t_{\mathrm{f}})$ are the position and
velocity vectors of the LMC $t_{\mathrm{f}}$ years ago.

These definitions require knowledge of the position and velocity of
the LMC over the past $350 \; \mathrm{Myr}$, for which we use an orbit
of the LMC around the MW taken from \citet{jethwa_magellanic_2016}, who rewound the
orbit of the LMC from the present position including the effect of
dynamical friction. The potential used by \citet{jethwa_magellanic_2016} for the MW is
an Navarro, Frenk \& White (1996, hereafter NFW) halo with mass
$10^{12}\;M_{\odot}$ and concentration $7.328$ and a Miyamoto \& Nagai
(1975) disk with radial scale-length $3 \; \mathrm{kpc}$ and vertical
scale-length $0.28 \; \mathrm{kpc}$ and a mass chosen to give the
solar circular velocity at the position of the sun. The LMC has an NFW
halo with mass $2\times10^{10}\;M_{\odot}$ and concentration $9.450$
and a Plummer bulge with scale-length $3 \; \mathrm{kpc}$ and a mass
chosen to satisfy the mass constraint of
\citet{van_der_marel_third-epoch_2014}.

\subsection{Milky Way potential}
\label{sec:potential}

For the MW, we use the potential from
\citet{kenyon_hypervelocity_2008}, which was optimised to reproduce
the galactic potential at both parsec scales near the MBH Sgr
A$^{\ast}$ and at scales of tens of kiloparsecs in the halo. This
potential consists of four components
\begin{equation}
\label{eq:potential}
\phi_{\mathrm{G}} = \phi_{\mathrm{bh}}+\phi_{\mathrm{b}}+\phi_{\mathrm{d}}+\phi_{\mathrm{h}}
\end{equation}
where
\begin{equation}
\label{eq:potentialbh}
\phi_{\mathrm{bh}}(r) = -GM_{\mathrm{bh}}/r
\end{equation}
is the Keplerian potential of the central MBH with $M_{\mathrm{bh}}=3.5\times10^6 \; \mathrm{M}_{\odot}$,
\begin{equation}
\label{eq:potentialb}
\phi_{\mathrm{b}}(r) = -GM_{\mathrm{b}}/(r+a_{\mathrm{b}}),
\end{equation}
is the Hernquist (1990) potential of the bulge with
$M_{\mathrm{b}}=3.76\times10^9\; \mathrm{M}_{\odot}$ and
$a_{\mathrm{b}} = 0.1 \; \mathrm{kpc}$,
\begin{equation}
\label{eq:potentiald}
\phi_{\mathrm{d}}(R,z) = -GM_{\mathrm{d}}/\sqrt{R^2+\left[a_{\mathrm{d}}+\left( z^2+b_{\mathrm{d}}^2 \right)^{1/2}\right]},
\end{equation}
is the Miyamoto-Nagai potential of the disk with
$M_{\mathrm{d}}=6\times10^{10} \; \mathrm{M}_{\odot}$,
$a_{\mathrm{d}}=2.75 \; \mathrm{kpc}$ and $b_{\mathrm{d}}=0.3 \;
\mathrm{kpc}$, and
\begin{equation}
\label{eq:potentialh}
\phi_{\mathrm{h}}(r) = -GM_{\mathrm{h}}\ln{\left(1+r/r_{\mathrm{h}}\right)}/r,
\end{equation}
is the NFW potential of the dark matter halo with $M_{\mathrm{h}} =
10^{12} \; \mathrm{M}_{\odot}$ and $r_{\mathrm{h}} = 20 \;
\mathrm{kpc}$. The values for each parameter are from \citet{kenyon_predicted_2014}.

\section{Discussion}
\label{sec:discussion}

\subsection{All sky density plots}
\label{sec:skydenplots}

\begin{figure}[t]
\includegraphics[scale=0.5,trim = 9mm 30mm 20mm 35mm, clip]{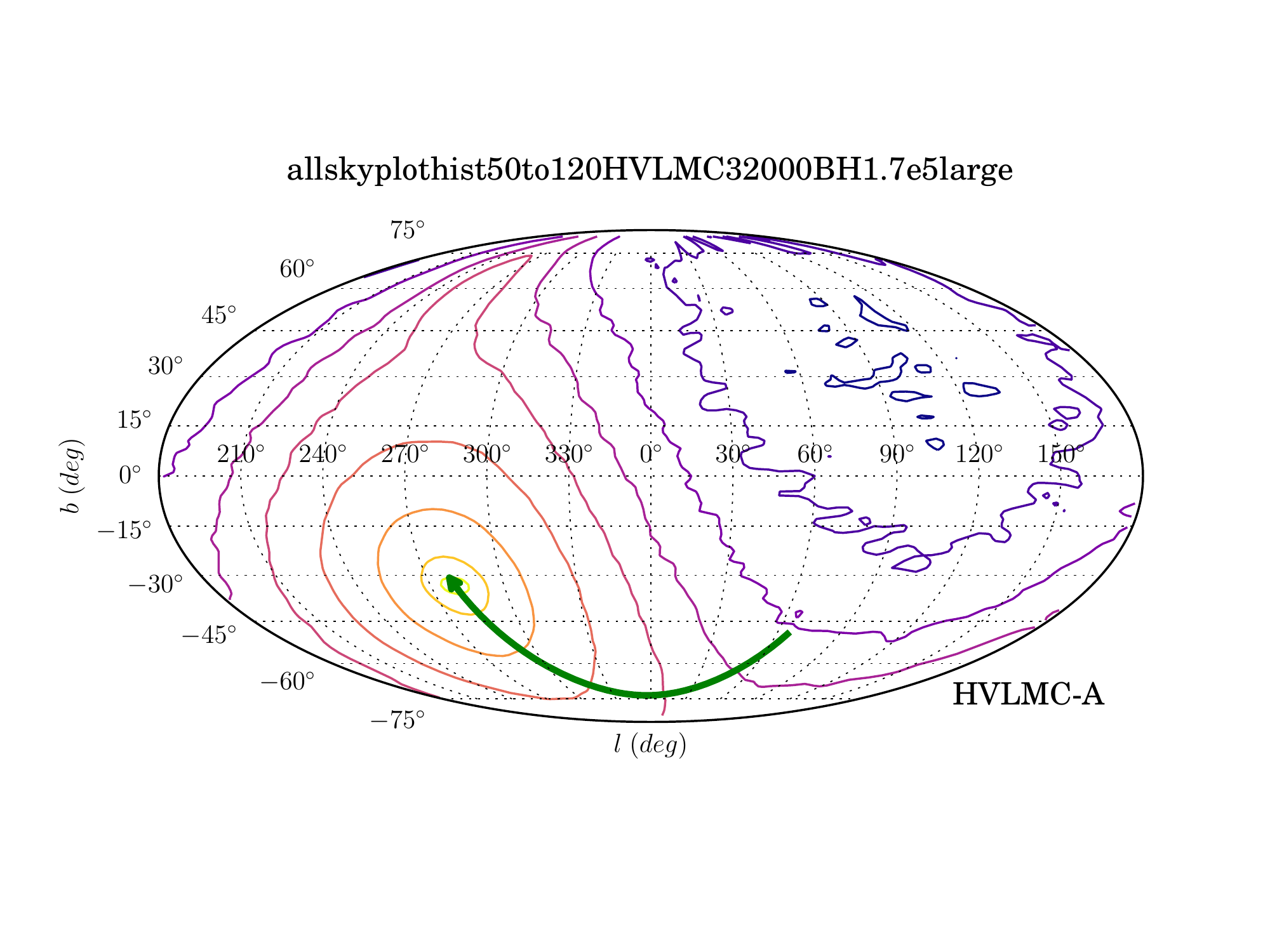}
\includegraphics[scale=0.5,trim = 9mm 30mm 20mm 35mm, clip]{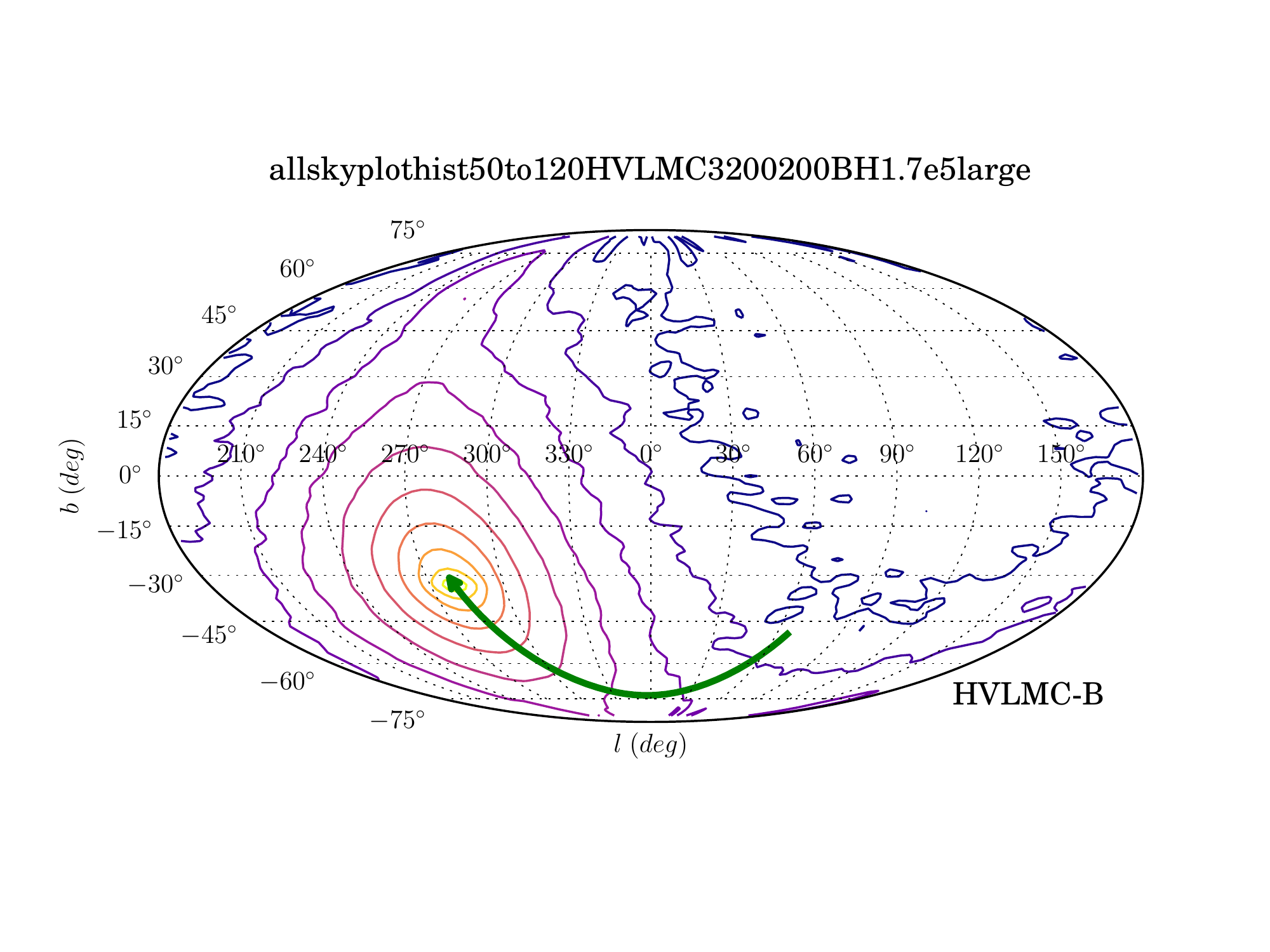}
\includegraphics[scale=0.5,trim = 9mm 30mm 20mm 35mm, clip]{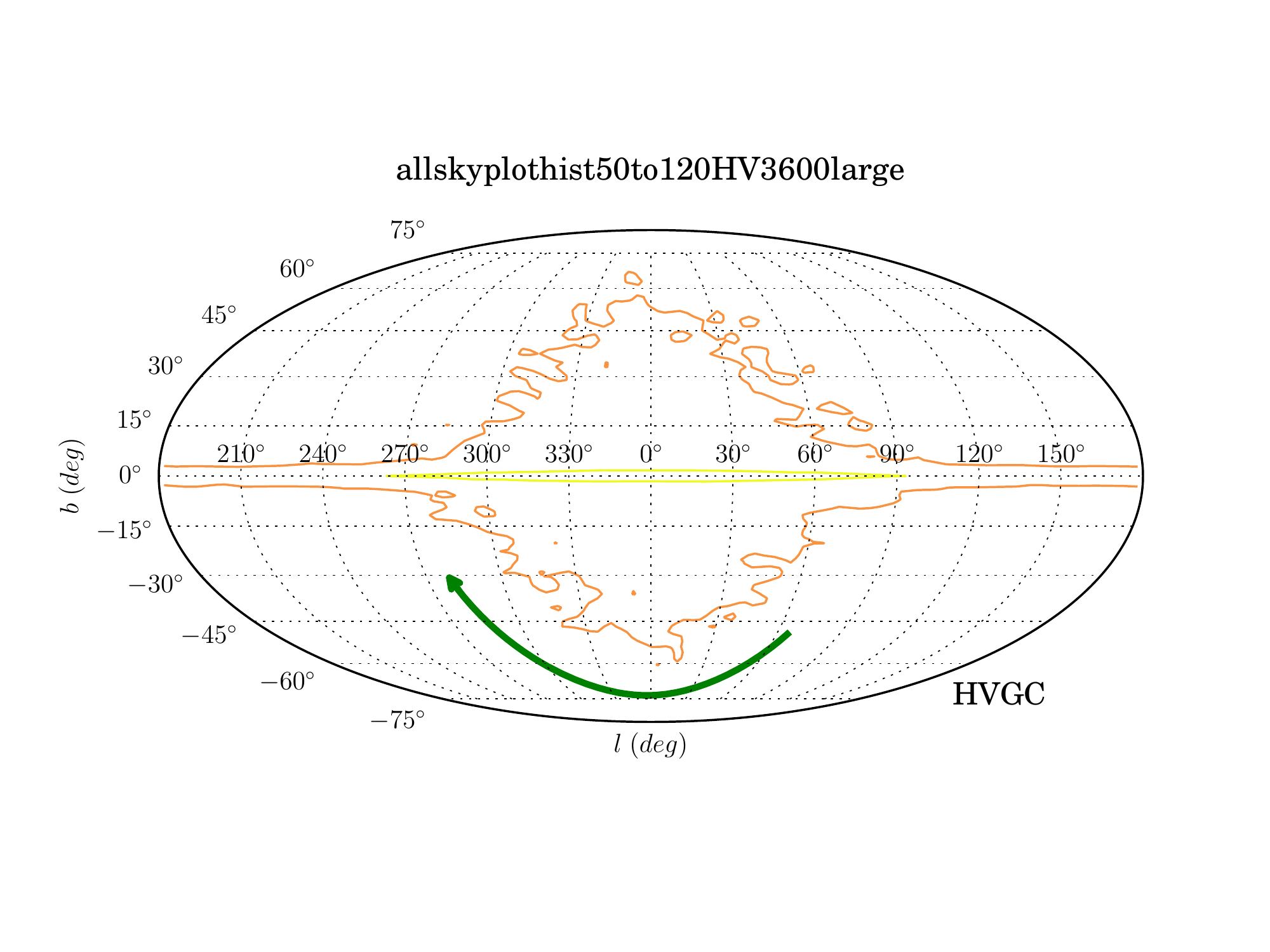}
\caption{Density contours in galactic coordinates for the stars from the three models with heliocentric distance $50 \leq d \leq 120 \; \mathrm{kpc} $.}
\label{fig:density}
\end{figure}

The density distribution of HVSs can be illustrated by an all-sky plot in galactic
coordinates, using the equal-area Mollweide projection. We consider only
stars with heliocentric distance $50<d<120\; \mathrm{kpc}$, since the distance of the LMC is $50.1\pm2.5
\; \mathrm{kpc}$ \citep{van_der_marel_third-epoch_2014} and this range covers the well-known clustering of
HVSs near the constellations of Leo and Sextans reported by
\citet{brown_anisotropic_2009}, however the kinematics in the distance bin $0<d<50\; \mathrm{kpc}$ are broadly similar. Figure \ref{fig:density} suggests that
the expected distribution of LMC HVSs on the sky is a
dipole, which contrasts with the monopole shown by HVSs from the
GC. Note that for the
HVGC model the first few contours aren't shown, since there are HVSs from the GC in our population across the entire sky.


We emphasise that the density contours are not normalised to the ejection rate from either MBH. \citet{brown_hypervelocity_2015} justifies an ejection rate of $10^{-4}\;\mathrm{yr}^{-1}$ for the MW MBH, but the ejection rate from a MBH at the centre of the LMC will depend significantly on the internal dynamics and star formation rate of the LMC as well as the mass of the MBH. We postpone these considerations to a companion paper, but speculate that the large contribution of the orbital velocity makes an observable signal of LMC HVSs plausible.

In the discovery paper of HVS11, \citet{brown_mmt_2009} commented that
that star was within $3.9^{\circ}$ of the Sextans dwarf galaxy while
the typical angular separation of HVSs from Local Group dwarf galaxies
is $10^{\circ}-20^{\circ}$. \citet{brown_mmt_2009} ruled out an
association with Sextans based on the $260 \; \mathrm{km} \; \mathrm{s}^{-1}$ relative velocity of HVS11 towards Sextans, however the initial instinct that
coincidence on the sky is a requirement for association is challenged
by Figure \ref{fig:density}. The LMC HVSs are distributed across a
wide region of the sky.

Figure \ref{fig:feature} highlights an intriguing extension of the
distribution that leads the LMC and is coincident with the clump
observed by \citet{brown_anisotropic_2009}. This is the only area of the sky which is both densely populated
in our LMC HVSs models and well-covered by HVS surveys, since almost
all of these surveys have covered solely the northern hemisphere,
partially due to the footprint of SDSS. The one HVS
plotted in Figure \ref{fig:feature} that lies in the southern
celestial hemisphere is HE 0437-5439 which was discovered by
\citet{edelmann_he_2005}, who noted that the flight time was longer
than the main sequence lifetime for the star and hence either it is a
blue straggler and was ejected as a binary from the GC or has its
origin in the LMC.

\citet{brown_anisotropic_2009} discusses previous attempts to explain
the anisotropy, including selection effects, contamination by runaway
companions to supernovae, tidal debris from disruption of a dwarf galaxy, temporary intermediate mass black hole companions
to the central MBH of the MW, and interactions between two stellar
discs near the central MBH which imprint their geometry on the ejected
HVSs. \citet{brown_hypervelocity_2015} states that the anisotropy is
still unexplained and that it will require a southern hemisphere
survey to resolve the issue.

\begin{figure}[h]
\includegraphics[scale=0.5,trim = 9mm 30mm 20mm 35mm, clip]{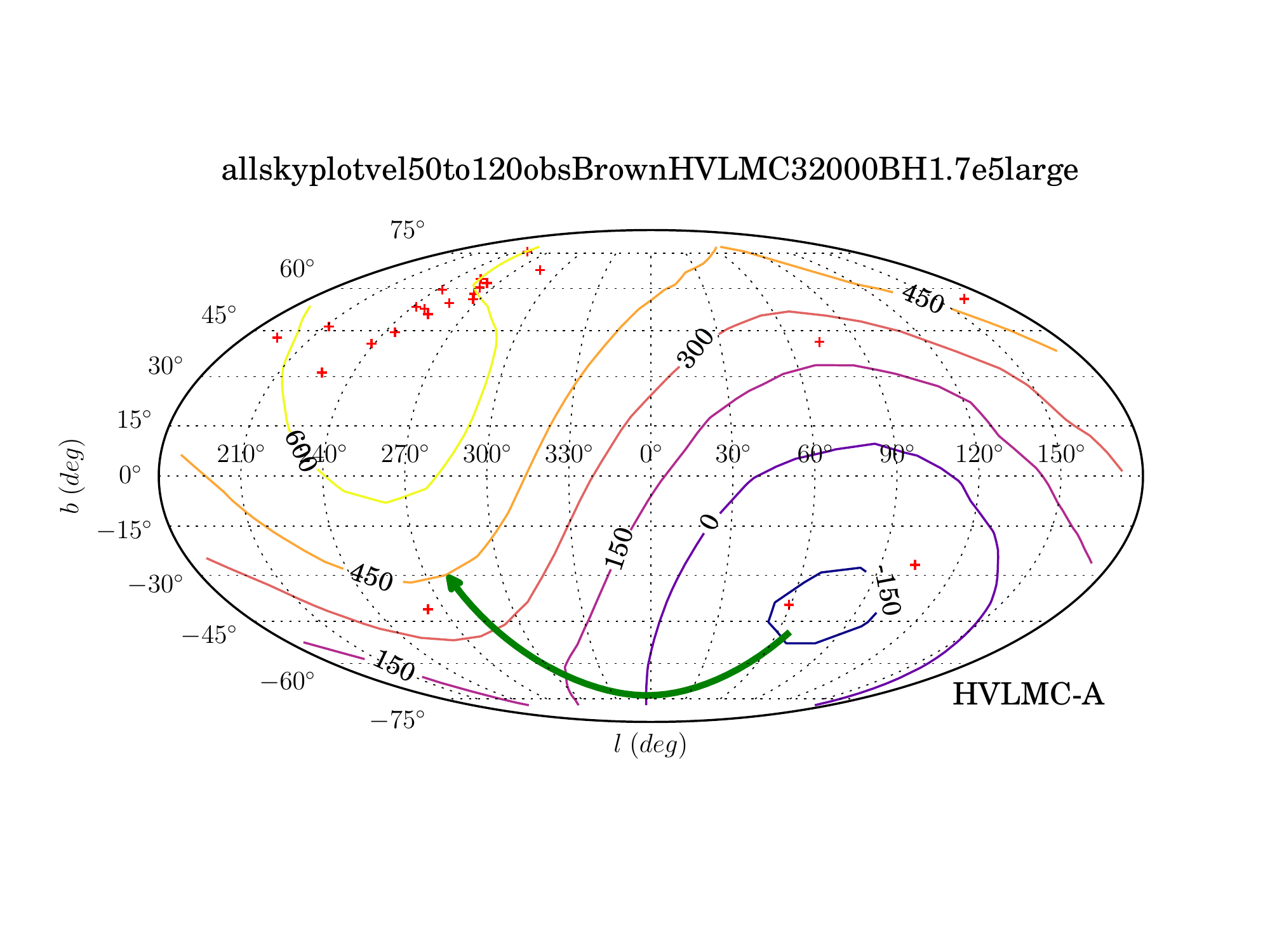}
\includegraphics[scale=0.5,trim = 9mm 30mm 20mm 35mm, clip]{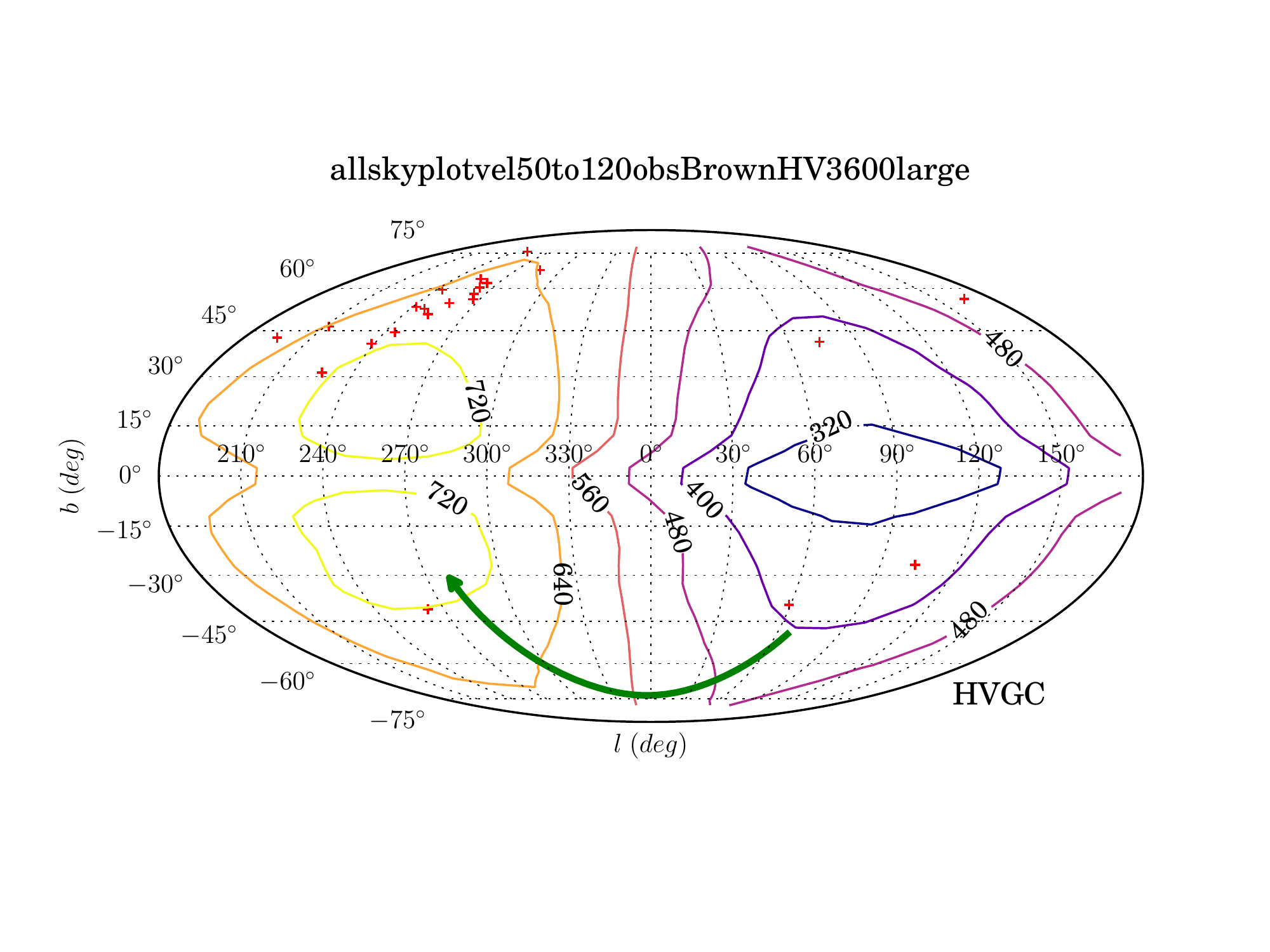}
\caption{Mean heliocentric radial velocity for stars at distance $50<d<120\; \mathrm{ kpc}$.}
\label{fig:vr}
\end{figure}


\begin{figure*}[t]
\includegraphics[scale=0.5,trim = 9mm 30mm 17mm 35mm, clip]{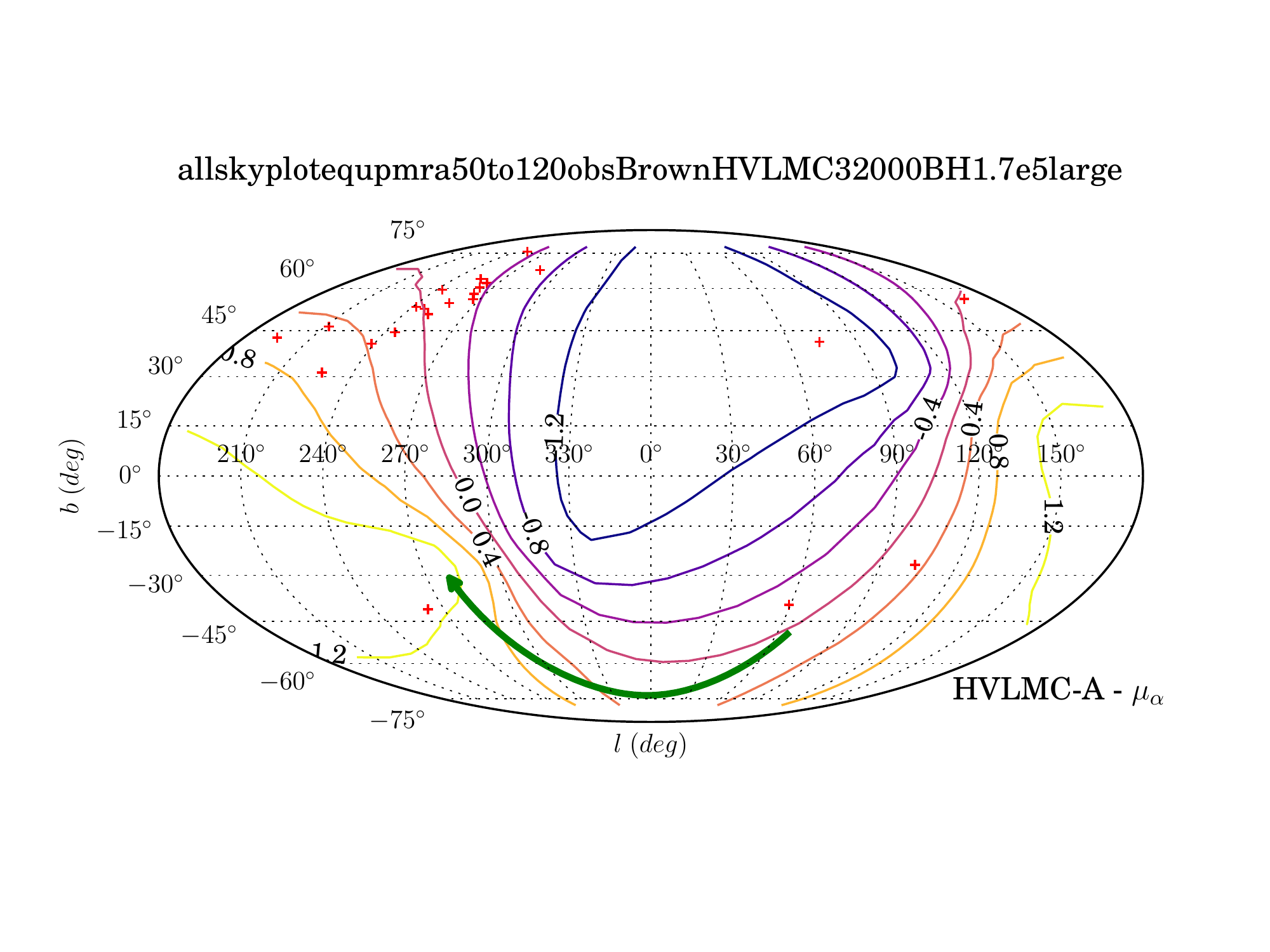}
\includegraphics[scale=0.5,trim = 9mm 30mm 17mm 35mm, clip]{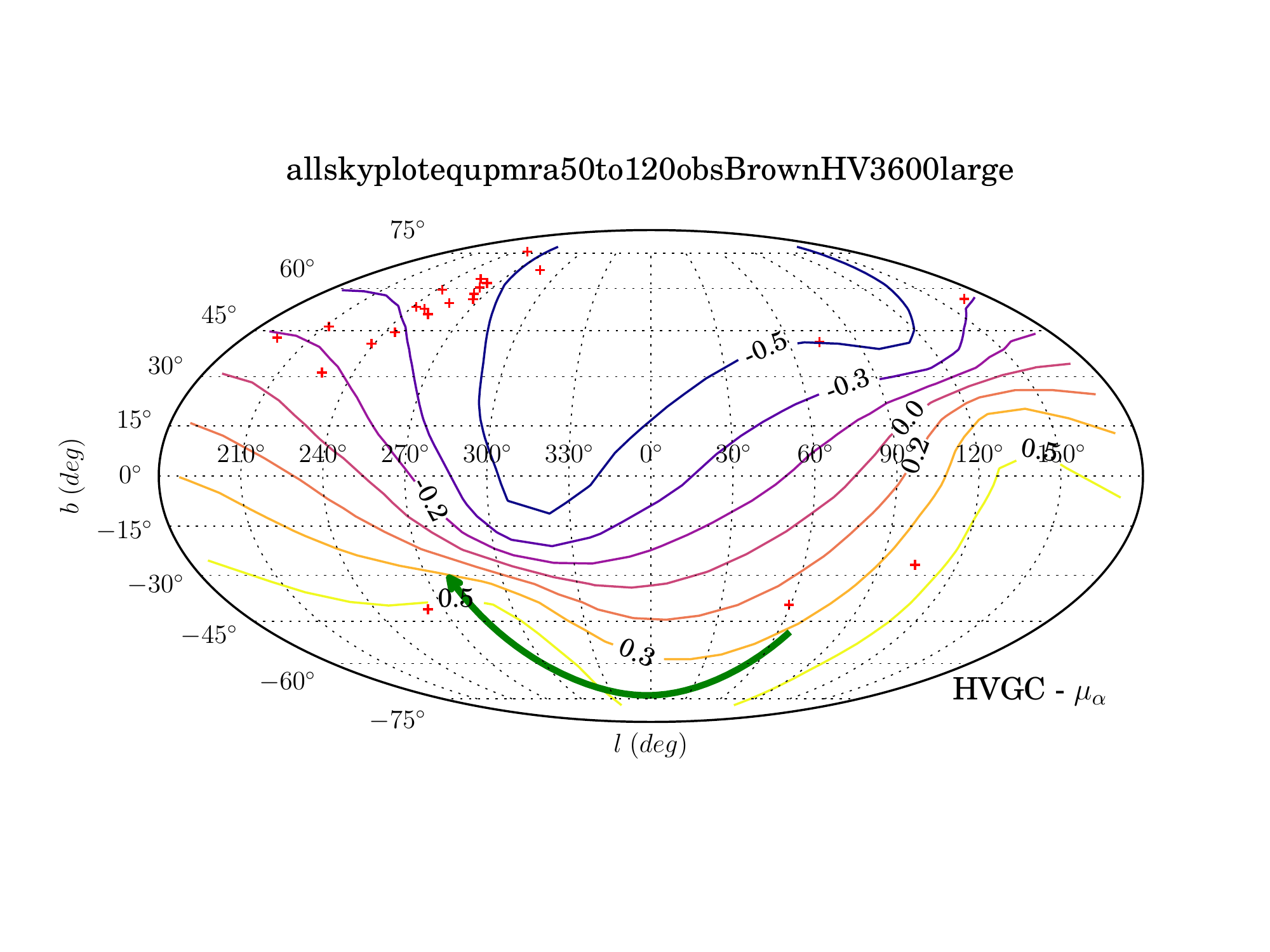}

\includegraphics[scale=0.5,trim = 9mm 30mm 17mm 35mm, clip]{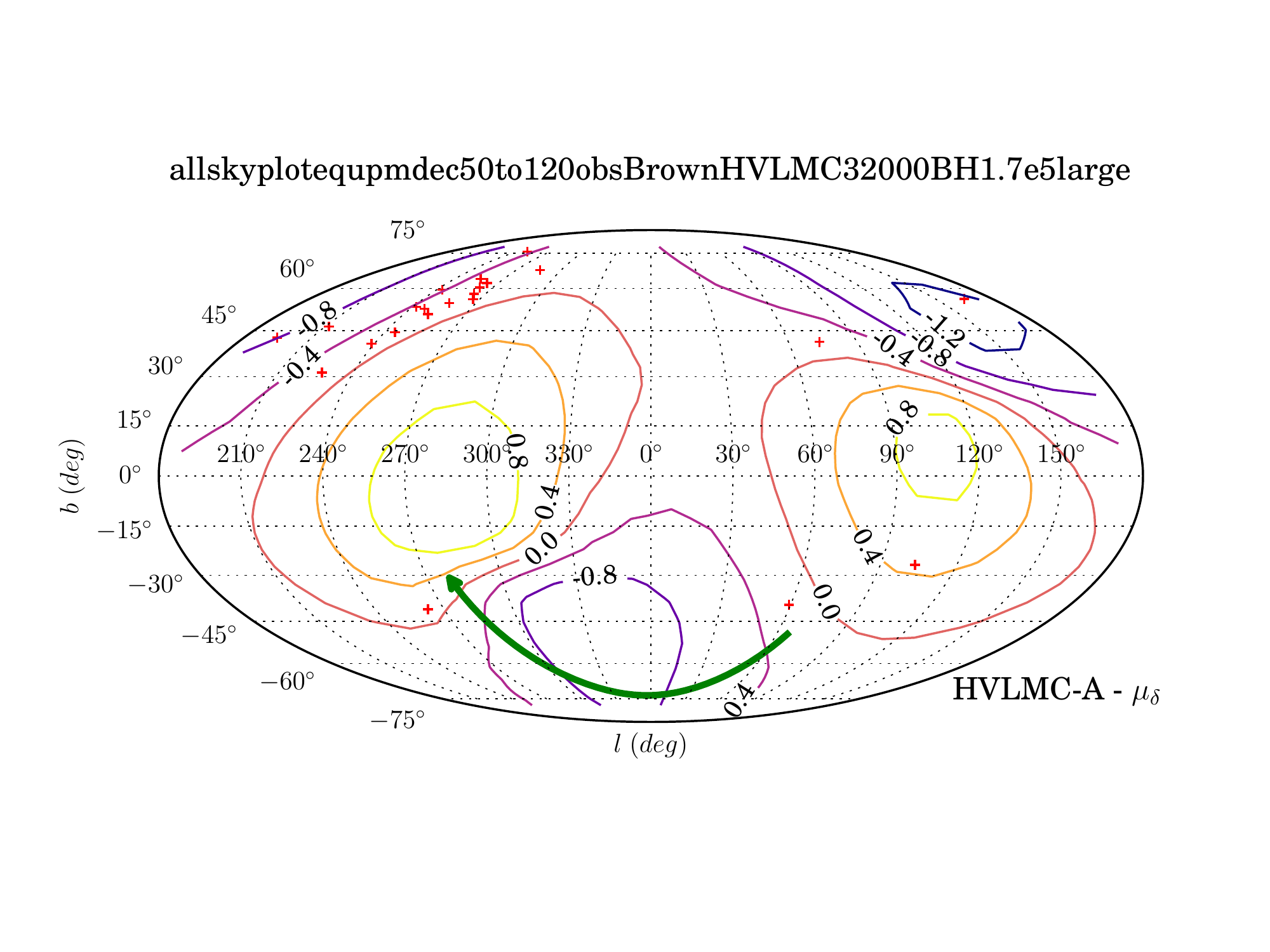}
\includegraphics[scale=0.5,trim = 9mm 30mm 17mm 35mm, clip]{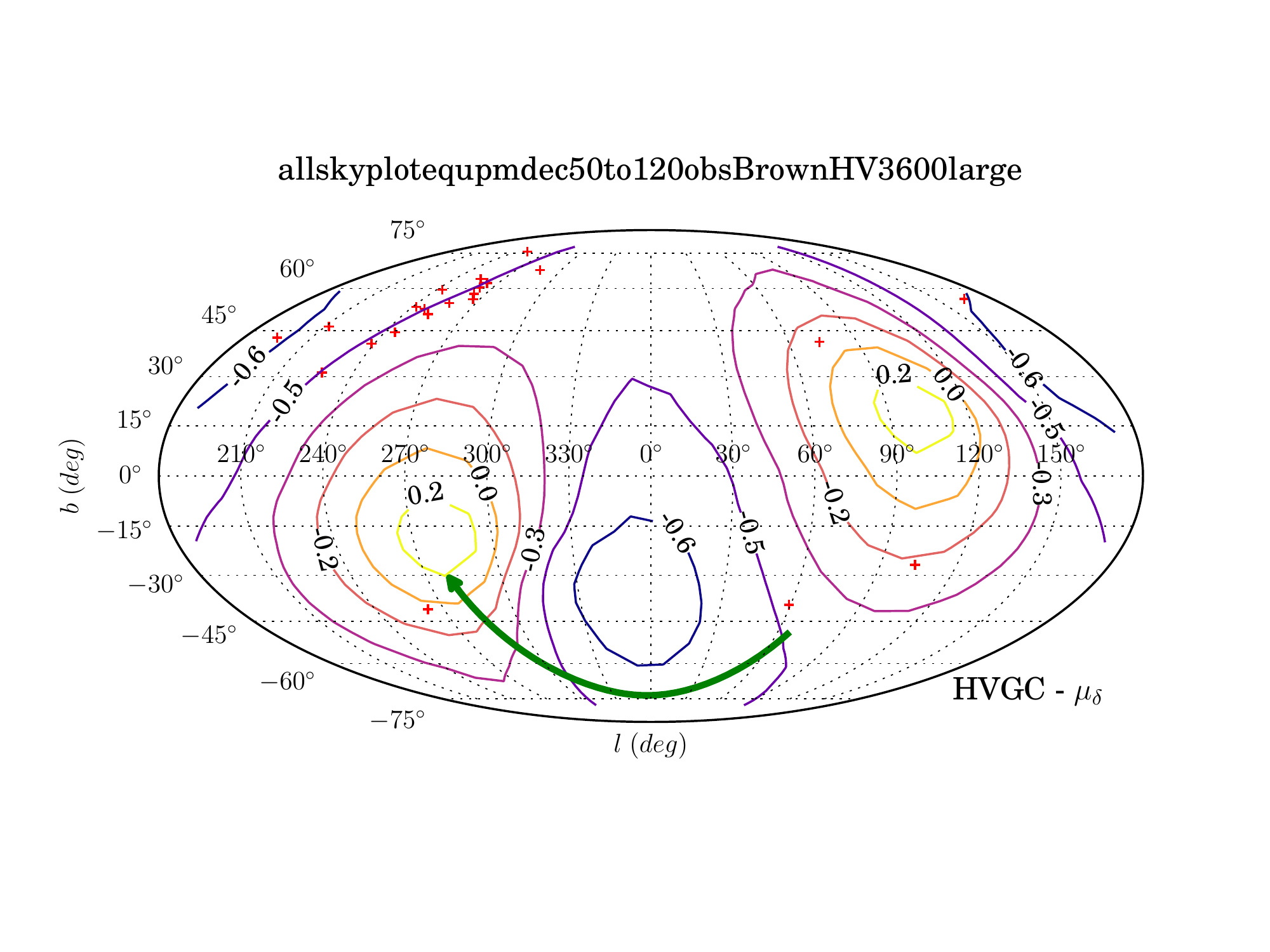}

\includegraphics[scale=0.5,trim = 9mm 30mm 17mm 35mm, clip]{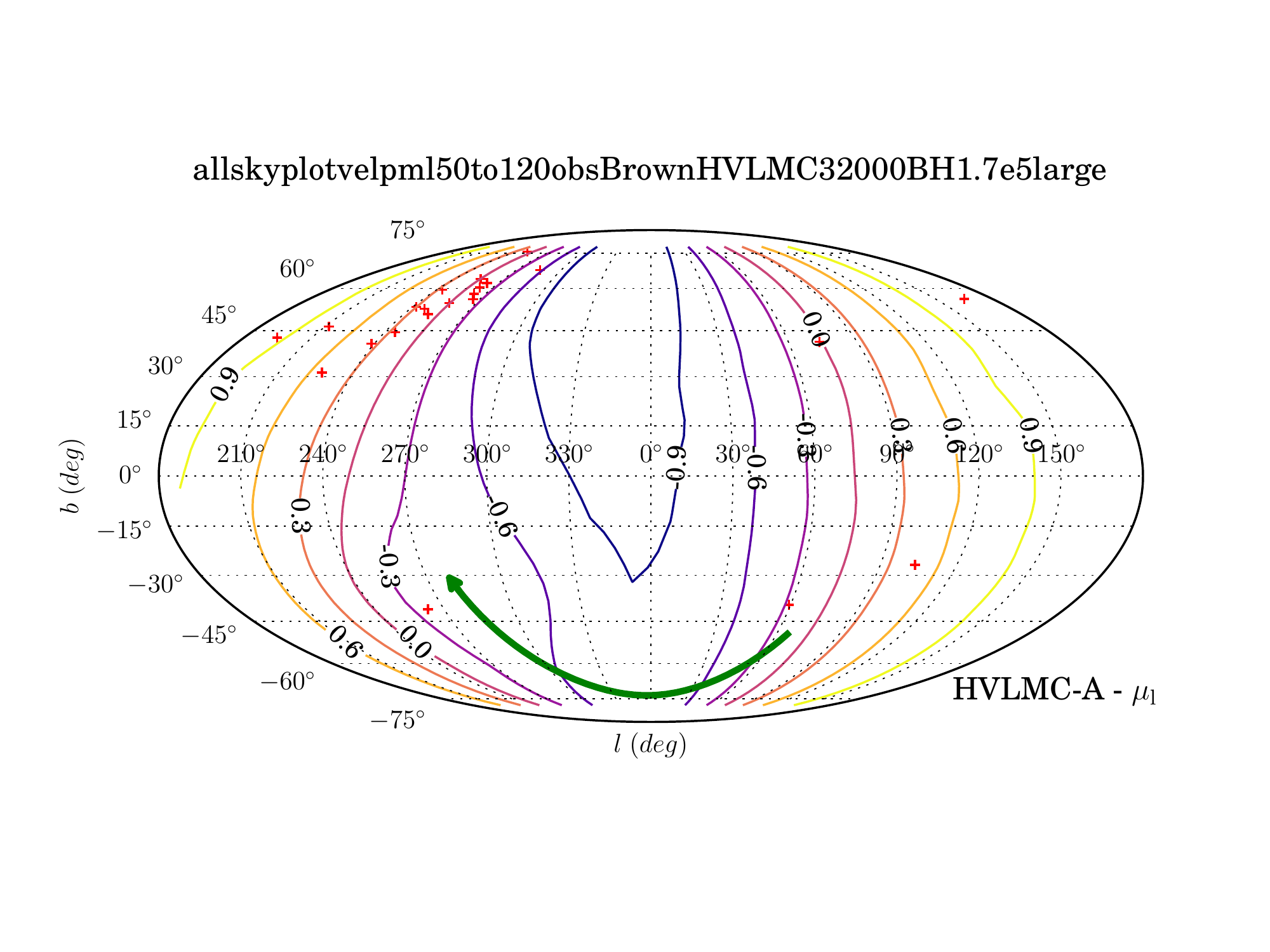}
\includegraphics[scale=0.5,trim = 9mm 30mm 17mm 35mm, clip]{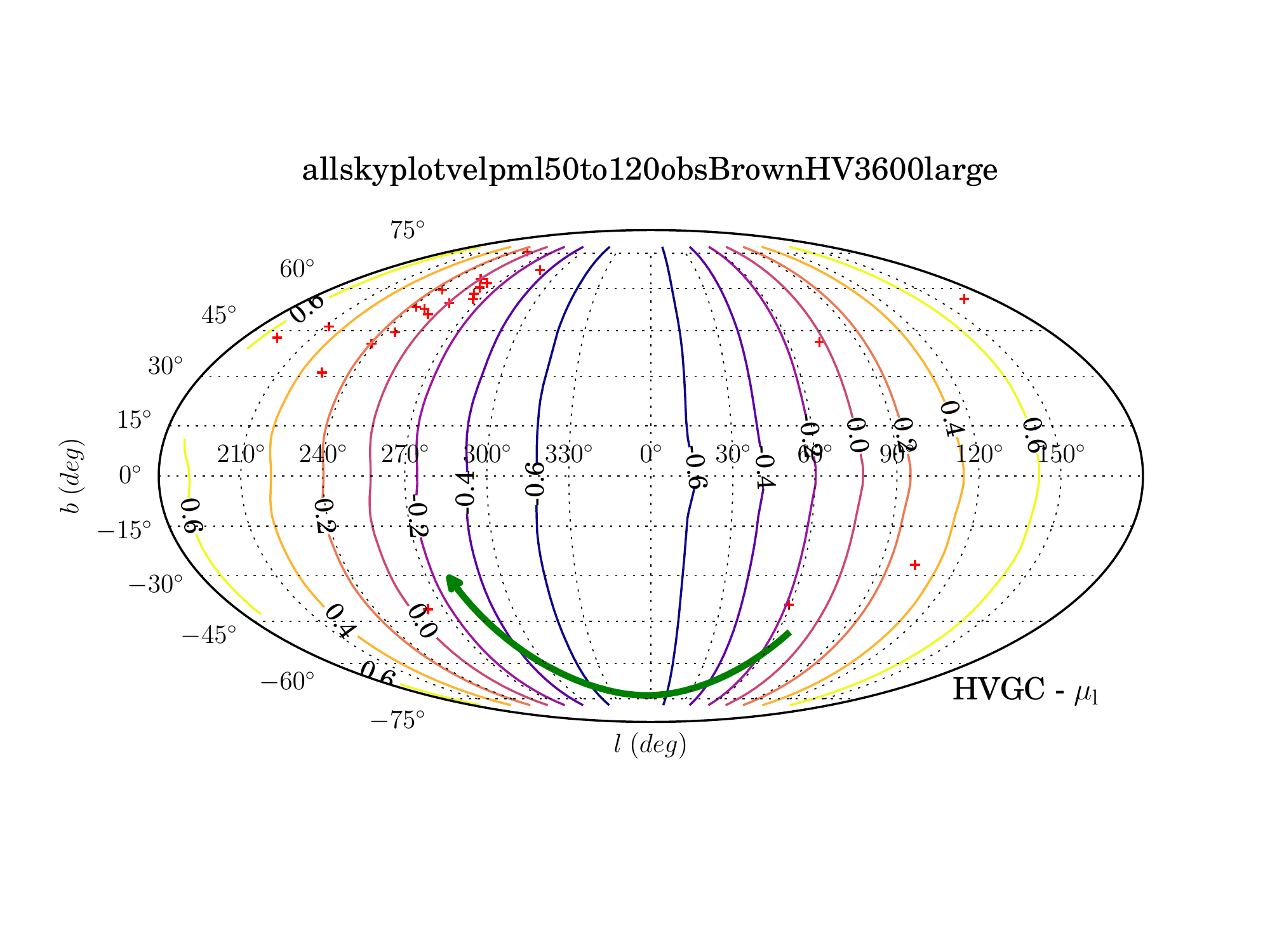}

\includegraphics[scale=0.5,trim = 9mm 30mm 17mm 35mm, clip]{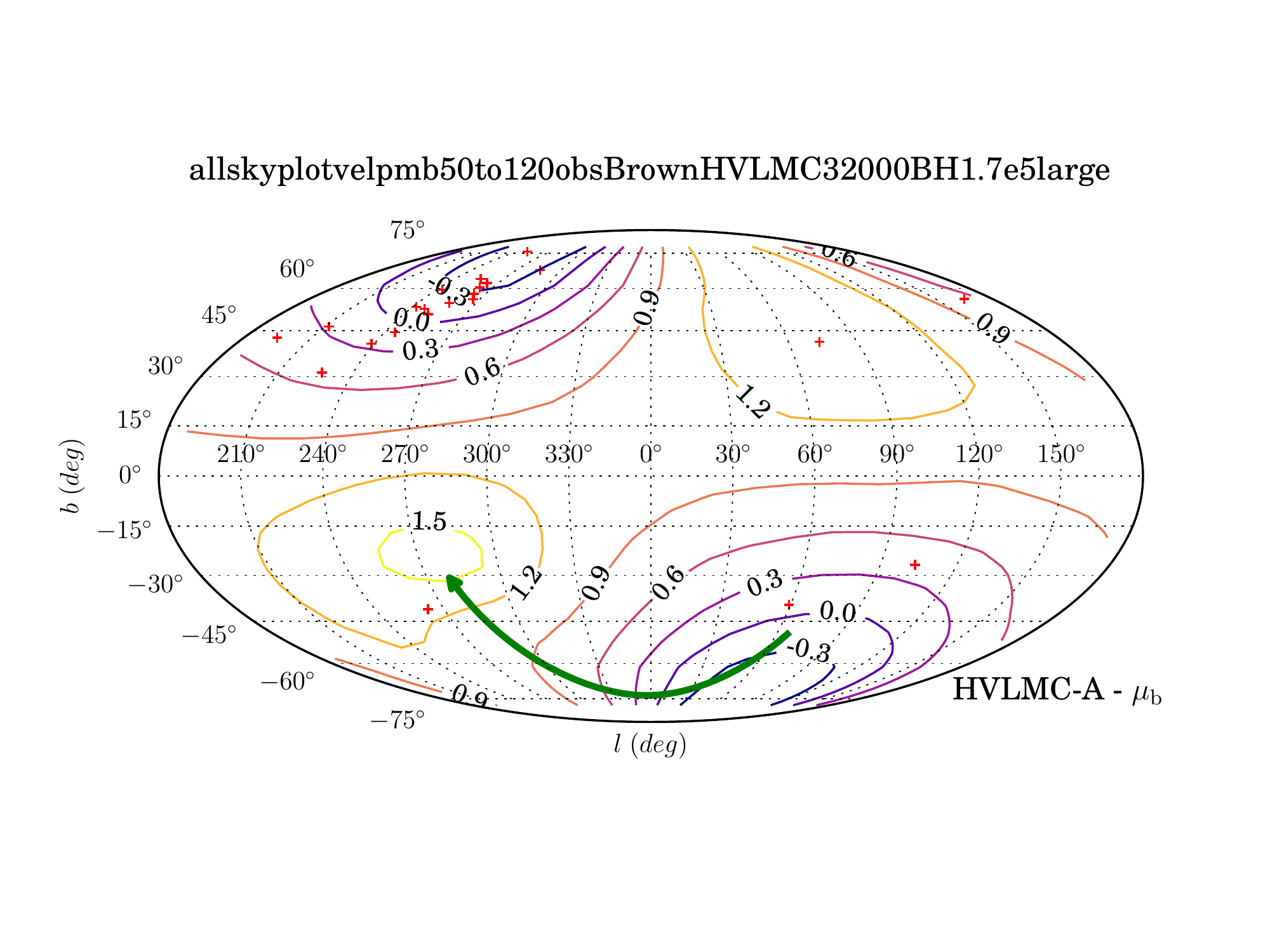}
\includegraphics[scale=0.5,trim = 9mm 30mm 17mm 35mm, clip]{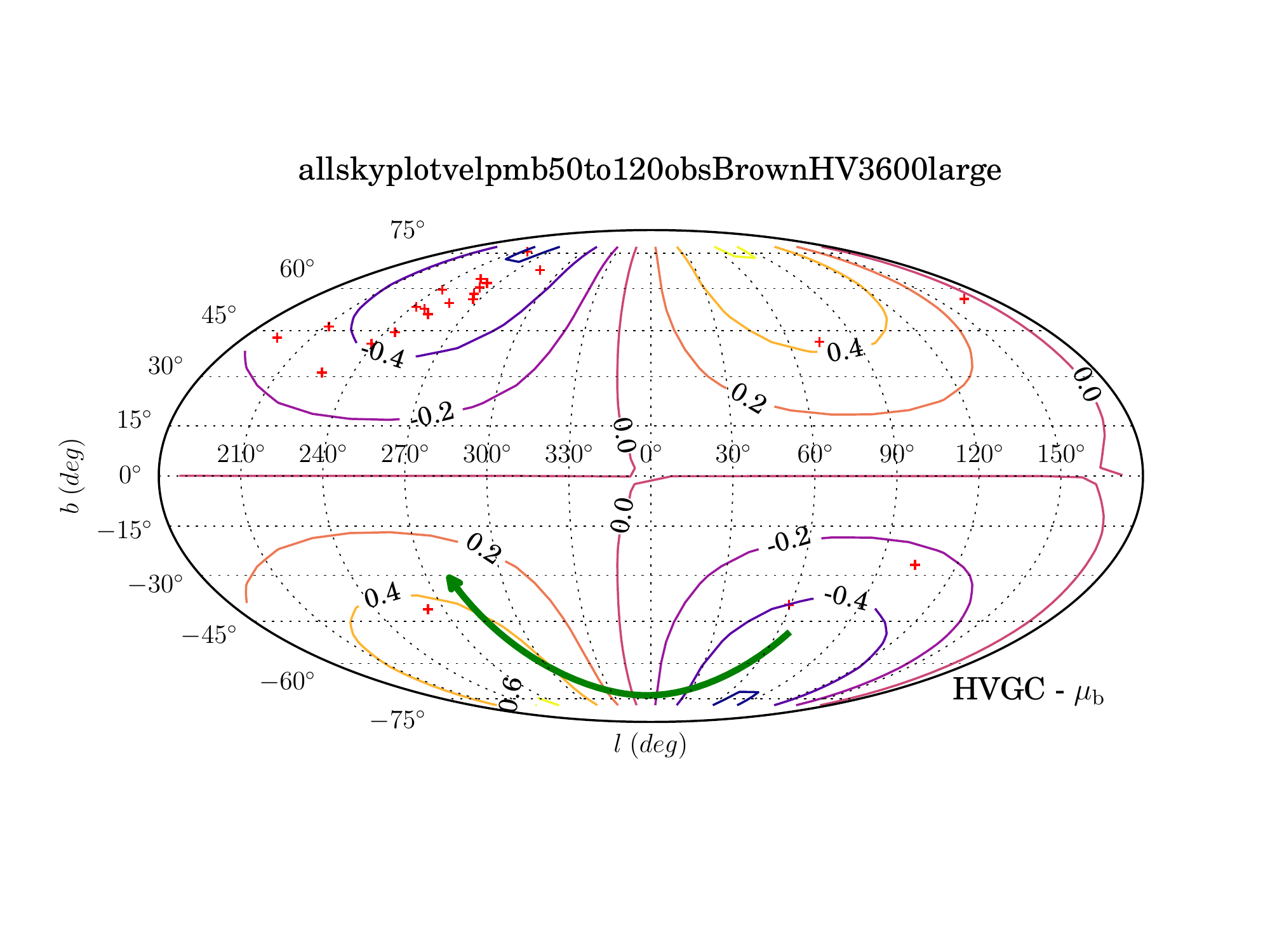}

\caption{Mean equatorial proper motions $(\mu_{\alpha},\mu_{\delta})$ and galactic proper motions $(\mu_{\mathrm{l}},\mu_{\mathrm{b}})$ for stars at distance $50<d<120\; \mathrm{ kpc}$.}
\label{fig:pm}
\end{figure*}

\subsection{All sky velocity plots}
\label{sec:skyvelplots}

In Figure \ref{fig:vr} we show contours of the mean heliocentric radial velocity in each bin, where the standard deviation in each plot is a couple $100 \; \mathrm{km} \; \mathrm{s}^{-1}$. The remaining two velocity dimensions are summarised in Figure \ref{fig:pm}, where we plot contours of the mean proper motion in each bin in both equatorial and galactic coordinates with a typical standard deviation of $0.5 \; \mathrm{mas}\;\mathrm{yr}^{-1}$. For the HVLMC-A model the contours demonstrate the imprint of the orbital velocity of the LMC in the kinematics of the HVSs, while for the HVGC stars the dominant effect is the solar rotation.

Considering Figures \ref{fig:vr} and \ref{fig:pm}, we see that at the locations on the sky of the currently observed B-type HVSs the velocity distributions of HVSs in the HVGC and HVLMC-A models are coincident. Since each model has a standard deviation of $0.5 \; \mathrm{mas}\;\mathrm{yr}^{-1}$ and the current proper motions for these stars have a typical error of $0.5 \; \mathrm{mas}\;\mathrm{yr}^{-1}$ \citep{brown_proper_2015}, it will not be possible to test our hypothesis using the current sample of HVSs. This situation should be resolved by Gaia, which will give proper motions for HVSs in regions on the sky where the two models differ by up to $0.7 \; \mathrm{mas}\;\mathrm{yr}^{-1}$. To illustrate this point, the first and third quartile ranges of the magnitude of the proper motions for stars at a distance $50<d<75 \; \mathrm{kpc}$ are $0.53 < \mu < 0.82 \; \mathrm{mas}\;\mathrm{yr}^{-1}$ in the HVGC model and $1.28 < \mu < 1.88\; \mathrm{mas}\;\mathrm{yr}^{-1}$ in the HVLMC-A model. Comparing these ranges to end-of-mission proper motion errors for Gaia\footnote{ \href{http://www.cosmos.esa.int/web/gaia/science-performance}{http://www.cosmos.esa.int/web/gaia/science-performance}} of around $450 \; \mu\mathrm{as}\;\mathrm{yr}^{-1}$ for B1V stars with apparent magnitude $V=20\;\mathrm{mag}$, corresponding to stars at about $100 \; \mathrm{kpc}$, it may be possible for Gaia to distinguish between the two populations.

\subsection{Spherical harmonics}
\label{sec:harmonics}

One method of quantifying the relative spatial distributions of HVSs
from the centre of the galaxy and the LMC is to consider the power in
each mode of the spherical harmonic power spectrum of the density.
For spherical harmonics defined by
\begin{equation}
\label{eq:ylm}
Y_\ell^m(\theta,\varphi)=\sqrt{\frac{(2\ell+1)}{4\pi}\frac{(\ell-m)!}{(\ell+m)!}}P_\ell^m(\cos\theta)e^{-im\varphi},
\end{equation}
where $\theta$ and $\varphi$ are the colatidudinal and longitudinal
coordinates, we can expand any function $f(\theta,\varphi)$ that is
square-integrable on the unit sphere as
\begin{equation}
\label{eq:fthetaphi}
f(\theta,\varphi)=\sum_{\ell=0}^{\infty}\sum_{m=-\ell}^{\ell}f_{\ell}^mY_{\ell}^m(\theta,\varphi).
\end{equation}
Using the orthonormality property of our chosen definition of the spherical harmonics we can then write
\begin{equation}
\label{eq:flmint}
f_{\ell}^m=\int_0^{2\pi}d\varphi \int_0^{\pi}d\theta \sin\theta \; f(\theta,\varphi)Y_{\ell}^{m\ast}(\theta,\varphi).
\end{equation}

\begin{figure}[t]
\includegraphics[scale=0.5,trim = 10mm 0mm 0mm 10mm, clip]{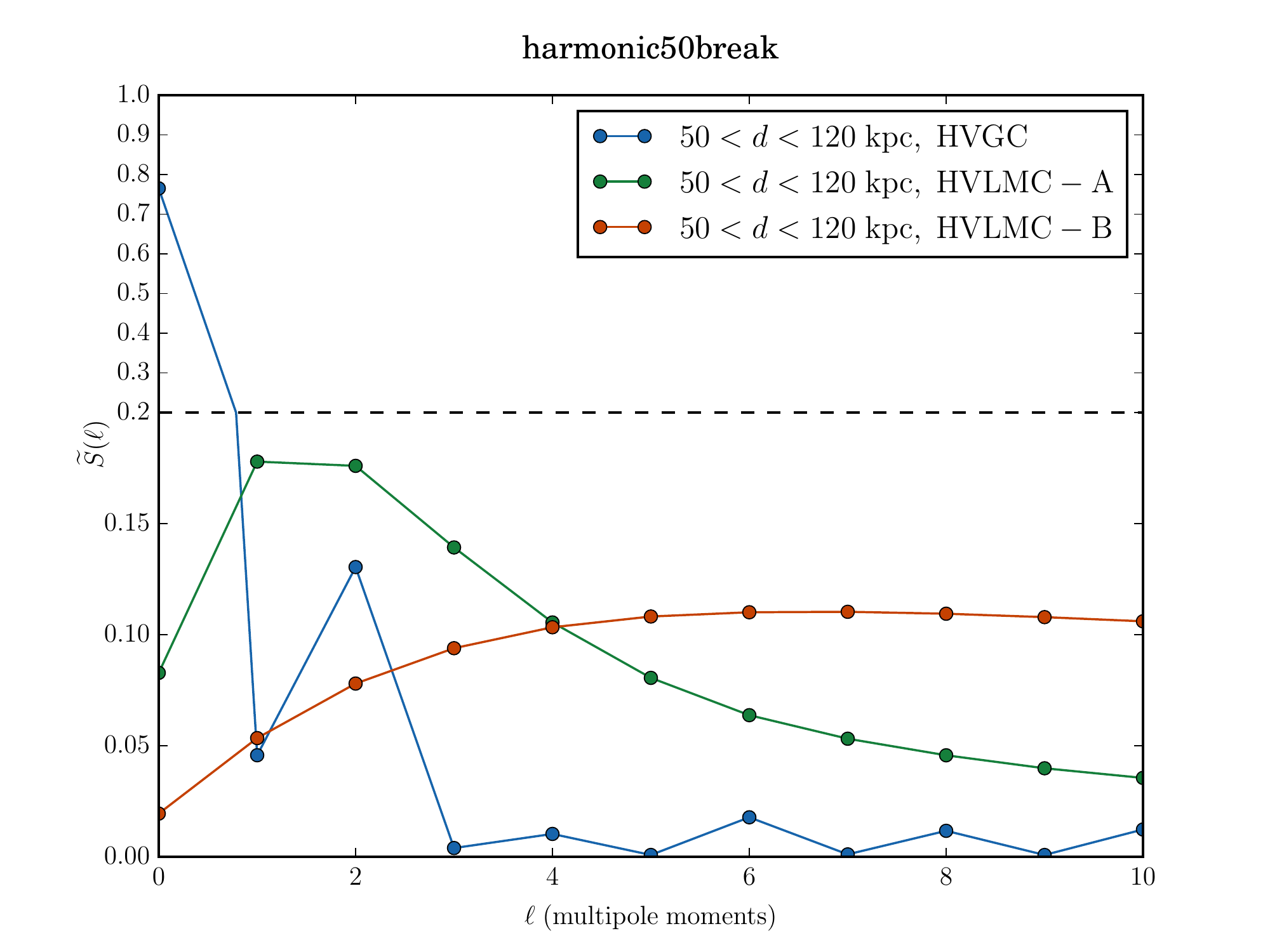}
\caption{Spherical harmonics for the distance bin $50<d<120 \; \mathrm{kpc}$. The power spectrum $S(\ell)$ has been normalised by the total power for each model to give $\widetilde{S}(\ell)$, since we are primarily interested in the relative power in each mode between models.}
\label{fig:harmonics}
\end{figure}

\noindent For our purposes, the function $f$ is a sum of delta
functions on the unit sphere at the locations $(l_i,b_i)$ of each of
the HVSs:
\begin{equation}
\label{eq:fbl}
f(b,l)=\sum_{i} \delta(b-b_i,l-l_i).
\end{equation}
Noting that $b=\frac{\pi}{2}-\theta,l=\varphi$, this sum of delta
functions transforms Equation \ref{eq:flmint} into a sum over the
HVSs,
\begin{equation}
\label{eq:flmsum}
f_{\ell}^m=\sum_{i} \cos{b_i} \; Y_{\ell}^{m\ast}\left(\frac{\pi}{2}-b_i,l_i\right).
\end{equation}
The angular power spectrum of $f$ is then given by
\begin{equation}
\label{eq:sff}
S(\ell)=4\pi\sum_{m=-\ell}^{\ell} \left| f_{\ell}^m \right|^2,
\end{equation}
and are plotted for our populations of HVSs in Figure \ref{fig:harmonics}.

Almost all of the power for the HVGC stars in this distance range is
in the monopole, which we expect as the highest velocity stars are
ejected along essentially straight lines from the GC, and
at $d>50 \; \mathrm{kpc}$ the difference between the the
observer being located at the sun or the GC is minimal. The HVLMC-A
model has a peak in power in the dipole/quadrupole, with a long tail
that is caused by the fast stars ejected ahead of the LMC and slow
stars lagging behind. The distinguishing features between the two
models are
\begin{enumerate}
\item a strong monopole in the HVGC model versus a strong dipole/quadrupole in the HVLMC-A model,
\item a large amount of power at high $\ell$ in the HVLMC-A model.
\end{enumerate}
The HVLMC-B model has a large section of the sky where there are no
HVSs, thus is not well-approximated by a dipole. The power is then
spread across a large number of modes since no one mode is a good
approximation. With the upcoming first data release of Gaia we may
soon be in a position to use spherical harmonic analysis to
distinguish between production mechanisms of HVSs.

\section{Conclusions}
\label{sec:conclusions}

Hypervelocity stars (HVSs) created in the Large Magellanic Cloud (LMC)
may contribute in a significant way to the sky distribution.  This may
provide a natural solution to the clustering of HVSs in the direction
of Leo and Sextans found by \citet{brown_anisotropic_2009}.  Uniquely,
this area of the sky is densely populated by LMC origin models and is
well-covered by current HVS surveys.


This hypothesis can shortly be tested by surveys in the southern
celestial hemisphere, such as SkyMapper and Gaia. If these surveys
fail to find a significant number of HVSs near the LMC, the model will
be falsified. A possible reason for failure is that the LMC does not
host a significant black hole. However, our choice of the Hills mechanism
as the production route was solely due to its proposed dominance in the
MW \citep{brown_hypervelocity_2015}. Other production routes which will be active in the disk of the LMC, such as
runaway companions of supernovae and three-body dynamical
interactions, can still result in high-velocity stars. Since the orbital velocity of
the LMC can provide $378 \; \mathrm{km} \; \mathrm{s}^{-1}$
\citep{van_der_marel_third-epoch_2014}, stars from these two
mechanisms only need to be ejected at the escape velocity of the LMC
($\gtrsim 100 \; \mathrm{km} \; \mathrm{s}^{-1}$) to be considered
anonymously high velocity stars.  The sky distributions of these
populations are under investigation in a companion paper.


The HVS candidate SDSS J121150.27+143716.2 was recently shown to be a
binary by \citet{nemeth_extremely_2016}, who concluded that the usual
production routes for HVSs cannot achieve a galactic rest frame
velocity of $571.3 \; \mathrm{km} \; \mathrm{s}^{-1}$ without
disrupting the binary and thus the binary is either an extreme halo
object or was accreted from the debris of a destroyed satellite
galaxy. While the kinematics of this candidate are inconsistent with
the LMC, we speculate that the addition of the orbital velocity of a
Local Group dwarf galaxy with the ejection velocity due to a standard
HVS production route could explain this HVS binary.

SkyMapper and Gaia will increase the quantity and quality of our HVS
sample across the entire sky and thus enable more sophisticated
analysis techniques, including our proposed spherical harmonic
analysis, which is capable of distinguishing between hypervelocity
populations in a quantitative way. Coincidence on the sky is not a
necessary requirement for association with a Local Group dwarf
galaxy. The satellites of the MW may well have imprinted distinctive
signatures on the distribution of HVSs right across the sky.


\section*{Acknowledgements}
DB thanks the STFC for a studentship. The orbit integration utilised
\texttt{galpy} \citep{bovy_galpy:_2015}. We thank Prashin Jethwa for
the use of his LMC orbit and Scott Kenyon for useful and prompt
answers to our queries. We further acknowledge the anonymous referee 
for comments that improved the clarity of our results.


\end{document}